\documentclass[%
 preprint,
 amsmath,amssymb,
 aps,pra
]{revtex4-1}

\usepackage{graphicx}
\usepackage{dcolumn}
\usepackage{bm}
\usepackage[caption=false]{subfig}

\begin{document}
\title{Theoretical Approach to Electroresistance in Ferroelectric Tunnel Junctions}

\author{Sou-Chi Chang}
\email{souchi@gatech.edu}
\author{Azad Naeemi}
\affiliation{School of Electrical and Computer Engineering, Georgia Institute of Technology, Atlanta, GA 30332 USA}
\author{Dmitri E. Nikonov}
\affiliation{Components Research, Intel Corporation, Hillsboro, OR 97124 USA}
\author{Alexei Gruverman}
\affiliation{Department of Physics and Astronomy, University of Nebraska, Lincoln, NE 68588 USA}

\begin{abstract}
In this paper, a theoretical approach, comprising the non-equilibrium Green's function method for electronic transport and Landau-Khalatnikov equation for electric polarization dynamics, is presented to describe polarization-dependent tunneling electroresistance (TER) in ferroelectric tunnel junctions. Using appropriate contact, interface, and ferroelectric parameters, measured current-voltage characteristic curves in both inorganic (Co/BaTiO$_{3}$/La$_{0.67}$Sr$_{0.33}$MnO$_{3}$) and organic (Au/PVDF/W) ferroelectric tunnel junctions can be well described by the proposed approach. Furthermore, under this theoretical framework, the controversy of opposite TER signs observed experimentally by different groups in Co/BaTiO$_{3}$/La$_{0.67}$Sr$_{0.33}$MnO$_{3}$ systems is addressed by considering the interface termination effects using the effective contact ratio, defined through the effective screening length and dielectric response at the metal/ferroelectric interfaces. Finally, our approach is extended to investigate the role of a CoO$_{x}$ buffer layer at the Co/BaTiO$_{3}$ interface in a ferroelectric tunnel memristor. It is shown that, to have a significant memristor behavior, not only the interface oxygen vacancies but also the CoO$_{x}$ layer thickness may vary with the applied bias. 
\end{abstract}

\maketitle
\section{Introduction}
Over the past four decades, the computing performance has been exponentially improved in a microchip because of doubled device density occuring approximately every two years according to the Moore's law \cite{Moore1965}. However, at the same time, as the complementary metal-oxide-semiconductor (CMOS) technology is down-scaled to the nanometer regime, the static power consumption plays a non-trivial role in total power dissipation due to a significant amount of leakage currents in memory and logic devices \cite{1250885}. As a consequence, recently, active research has also been underway in pursuit of low-power and non-volatile memory and logic circuits in the beyond-CMOS technologies \cite{7076743}, and the major advantages of the non-volatility in the microprocessor potentially are (i) the system speed improvement by eliminating the need of transferring data between volatile power-starving memories (i.e. static and dynamic random-access memories) and external non-volatile storage (i.e. hard disk drive) as well as (ii) the energy efficiency enhancement by removing the static power consumption.

Among many emerging non-volatile memory technologies, ferroelectric (FE) devices based on quantum-mechanical tunneling, known as ferroelectric tunnel junctions (FTJs), have attracted significant attention due to the extremely high ON/OFF ratio, very low write power, and non-destructive read \cite{Ionescu2012}. The concept of an FTJ has been demonstrated experimentally \cite{doi:10.1021/nl901754t,:/content/aip/journal/apl/83/22/10.1063/1.1627944,Chanthbouala2012,Garcia2014,Tian2016} thanks to improved technologies in fabricating high quality ultra-thin FE films by pulsed laser deposition or off-axis sputtering, which push the critical thickness of ferroelectricity down to a few unit cells \cite{:/content/aip/journal/apl/84/25/10.1063/1.1765742,:/content/aip/journal/apl/86/10/10.1063/1.1880443,PhysRevLett.89.067601,Fong1650,PhysRevB.72.020101}. Moreover, over the past decade, FE fabrication technologies have become mature and compatible to the back-end CMOS process \cite{929749}, and therefore FTJ-CMOS circuits with additional microchip functionality may become a reality in the near future. 

In an FTJ, the switching of resistance, also known as tunneling electroresistance (TER) effect, is achieved by the polarization reversal in the FE barrier via applied voltage. The TER effect is fundamentally different from other resistive switching mechanisms such as the formation of conductive filaments within a metal-oxide insulator in an atomic switch \cite{5557741}, the oxygen-vacancy-assisted conduction in a resistive random-access memory (RAM) \cite{5607274}, and the magnetization-dependent tunneling in a magnetic tunnel junction (MTJ) \cite{1200123}. In particular, unlike tunneling magnetoresistance (TMR) in the MTJ, which is typically only a few hundred percent \cite{Yuasa2004,Parkin2004,:/content/aip/journal/apl/93/8/10.1063/1.2976435}, TER in an FTJ can easily reach $10^{5}\%$ \cite{Garcia2014}, offering a much more reliable $\it{read}$ mechanism for the stored memory bits. While significant TER is achieved in FTJs, there still exists a controversy in TER signs, particularly for Co/BaTiO$_{3}$/La$_{0.67}$Sr$_{0.33}$MnO$_{3}$ (Co/BTO/LSMO) systems \cite{doi:10.1021/nl302912t,Chanthbouala2012}; that is, TER signs observed experimentally from different groups are completely opposite. Note that the term "TER sign" is introduced here to specify the relation between the electric polarization direction and the resistance state. The TER sign is defined as "$+$" (positive) and "$-$" (negative) when the low (ON) resistance state is produced by the polarization pointing to the top and the bottom electrodes, respectively. Recent experimental work shows that these opposite TER signs can be attributed to the dead layers induced by either TiO$_{2}$ or BaO termination at the Co/BTO interface \cite{ADFM:ADFM201500371}.

In addition to the promising progress in the FTJ experiments, lots of theoretical efforts have also been made in predicting or understanding TER in an FTJ. Inspired by the polar switch concept proposed by Leo Esaki in $1971$ \cite{leo}, the giant TER was predicted near the zero bias based on electron direct tunneling \cite{Tsymbal181,PhysRevLett.94.246802}. Using a similar model, enhanced TER by inserting a non-polar dielectric layer at the metal/FE interface was also predicted near the equilibrium \cite{:/content/aip/journal/apl/95/5/10.1063/1.3195075}. Furthermore, going beyond the equilibrium, polarization-dependent TER was predicted to be based either solely on direct tunneling \cite{PhysRevB.72.125341} or on combination of several transport mechanisms including direct tunneling, Fowler-Nordheim tunneling, and thermionic emission \cite{PhysRevB.82.134105}. Nevertheless, works on polarization-dependent TER were mainly based on the analytical models derived from the Wenzel-Kramer-Brillouin (WKB) approximation and did not include a realistic FE hysteresis loop. More importantly, most of the theoretical approaches describe the experimental data in the low-voltage range; so far, none of them has provided quantitative comparisons with current-voltage ($I$-$V$) characteristics measured from a full FE hysteresis sweep, which is extremely important in designing FTJs as memory elements, where both $\it{read}$ and $\it{write}$ operations need to be well-described. This paper presents a comprehensive approach to (i) describe the experimentally measured $I$-$V$ relations for various types of FTJs, and (ii) to explain the discrepancy in the TER signs observed experimentally by different groups in the Co/BTO/LSMO layered structures. The developed approach includes the non-equilibrium Green’s function (NEGF) method for electronic transport under different bias conditions \cite{Supriyo_2005} and the thermodynamics-based Landau-Khalatnikov equation for a complete ferroelectric hysteresis loop.

\begin{figure}[h!]
\begin{center}
\includegraphics[width=2.5in]
{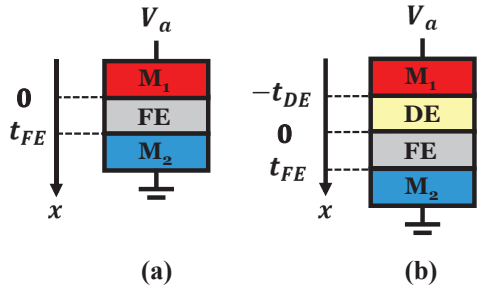}
\caption{Schematics of FTJs in the (a) absence and (b) presence of a non-polar dielectric (DE) layer between the ferroelectric (FE) and metal electrode. M$_{1}$ and M$_{2}$ are top and bottom metal electrodes, respectively.}
\label{fig1}
\end{center}
\end{figure}

An FTJ structure is shown in Fig. \ref{fig1}(a), where the device is composed of an FE thin film sandwiched between two metal electrodes. In this work, TER is assumed to be induced by band structure modifications through the electrostatic effect due to polarization reversal (Fig. \ref{fig3}). Moreover, to explore the role of a CoO$_{x}$ buffer layer in the Co/BTO/LSMO systems, reported to be an inevitable by-product while depositing the metallic electrode \cite{doi:10.1021/nl302912t}, an FTJ structure with a non-polar DE layer at the metal/FE interface is also considered as shown in Fig. \ref{fig1}(b).

The rest of this paper is organized as follows. In Section II, mathematical details of the proposed approach for TER in an FTJ is presented. In Section III, using this theoretical model, good agreement with the experimental $I$-$V$ characteristics is shown for various FTJs, and the discrepancy in the reported TER sign in Co/BTO/LSMO systems is explained by introducing the termination effect using the effective contact ratios. Also, the model is extended to investigate the role of a CoO$_{x}$ buffer layer in an FE memristor. Section IV concludes the paper.

\section{Theoretical Formalism}

\subsection{FTJ without Non-polar Dielectric}
\begin{figure}[h!]
\begin{center}
\includegraphics[width=3.6in]
{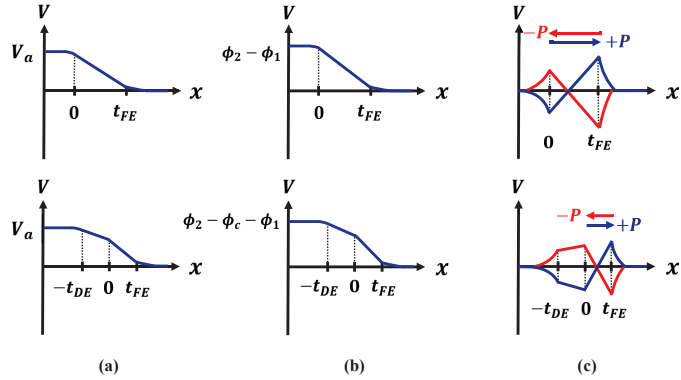}
\caption{Schematics of electrostatic potential profiles due to (a) applied electric field, (b) built-in field, and (c) depolarization field for FTJs with (bottom panel) and without (top panel) a non-polar DE layer between the FE and top metal electrode.}
\label{fig2}
\end{center}
\end{figure}

To describe the polarization-dependent TER in an FTJ, the energy band diagram under the effects of the applied electric field, built-in field, and depolarization field is considered. In this work, the applied electric field is generated by a bias voltage across an FTJ, the built-in field is mainly due to the work function difference between layered materials \cite{PhysRevLett.98.207601,PhysRevB.77.174111,PhysRevB.88.024106}, and the depolarization field is induced by the incomplete screening of the FE bound charge. Figs. \ref{fig2}(a), (b), and (c) illustrate electrostatic potential profiles induced by the applied electric field, built-in field, and depolarization field for FTJs in the presence and absence of a non-polar DE layer, respectively. Mathematically, for an FTJ without a non-polar DE layer, it is assumed that the potential profiles within metals ($V_{M1}$ and $V_{M2}$) follow the Thomas-Fermi expression \cite{,PhysRevApplied.4.044014} and are given as (see Appendix A for detailed derivations)
\begin{eqnarray}
V_{M1}\left(x\right) = \frac{-\rho_{s}\lambda_{1}}{\epsilon_{1}\epsilon_{0}}e^{\frac{x}{\lambda_{1}}},
\label{eq1} \\
V_{M2}\left(x\right) = \frac{\rho_{s}\lambda_{2}}{\epsilon_{2}\epsilon_{0}}e^{\frac{-\left(x-t_{FE}\right)}{\lambda_{2}}},
\label{eq2}
\end{eqnarray}
where $\rho_{s}$ is the screening charge density at the FE/metal interfaces (C/m$^{2}$), $\lambda_{1}$ and $\lambda_{2}$ are effecitve screening lengths of top and bottom FE/metal interfaces, respectively, $\epsilon_{1}$ and $\epsilon_{2}$ are relative dielectric constants of top and bottom FE/metal interfaces, respectively, and $\epsilon_{0}$ is the vacuum dielectric constant. Note that the imperfect screening here is described by both effective screening length and dielectric constant, rather than Thomas-Fermi one, since it is generally accepted that the imperfect screening is determined not only by the metal, but also by the FE thin film and the specific interface geometry \cite{Junquera:2008-11-01T00:00:00:1546-1955:2071}. As a result, from Eqs. \ref{eq1} and \ref{eq2}, the potential drop in top and bottom electrodes are $\frac{\rho_{s}\lambda_{1}}{\epsilon_{1}\epsilon_{0}}$ and $\frac{\rho_{s}\lambda_{2}}{\epsilon_{2}\epsilon_{0}}$, respectively. By assuming that the electric displacement is continuous throughout the FTJ, the following equation is held.
\begin{eqnarray}
\rho_{s}=\epsilon_{0}E_{FE}+P, \label{eq3}
\end{eqnarray}
where $P$ is the electric polarization of the FE and $E_{FE}$ is the total electric field across the FE. Furthermore, due to the fact that the potential drop induced by the applied bias and built-in field has to be completely shared by both metal electrodes and the FE, the following equation is satisfied.
\begin{eqnarray}
\frac{\rho_{s}\lambda_{1}}{\epsilon_{1}\epsilon_{0}}+\frac{\rho_{s}\lambda_{2}}{\epsilon_{2}\epsilon_{0}}+E_{FE}t_{FE} = V_{a}+V_{bi}, \label{eq4}
\end{eqnarray}
where $V_{a}$ is the applied voltage and $V_{bi}$ is the voltage drop due to the built-in field, defined as $\frac{\phi_{2}-\phi_{1}}{e}$ with $\phi_{1}$ and $\phi_{2}$ being conduction band discontinuities at the top and bottom FE/metal interfaces, respectively, and $e$ being the elementary charge. From Eqs. \ref{eq3} and \ref{eq4}, the total electric field across the FE is given as
\begin{eqnarray}
E_{FE}=\frac{V_{a}+V_{bi}-P\left(\frac{\lambda_{1}}{\epsilon_{1}\epsilon_{0}}+\frac{\lambda_{2}}{\epsilon_{2}\epsilon_{0}}\right)}{t_{FE}+\frac{\lambda_{1}}{\epsilon_{1}}+\frac{\lambda_{2}}{\epsilon_{2}}}. \label{eq5}
\end{eqnarray}
Note that the depolarization field, $E_{dep}$, is obtained by canceling the built-in field with the applied bias ($V_{a}+V_{bi}=0$) and given as
\begin{eqnarray}
E_{dep}=\frac{-P\left(\frac{\lambda_{1}}{\epsilon_{1}\epsilon_{0}}+\frac{\lambda_{2}}{\epsilon_{2}\epsilon_{0}}\right)}{t_{FE}+\frac{\lambda_{1}}{\epsilon_{1}}+\frac{\lambda_{2}}{\epsilon_{2}}}. \label{eq6}
\end{eqnarray}
By replacing $E_{FE}$ in Eq. \ref{eq3} with Eq. \ref{eq6}, the screening charge density induced simply by the FE bound charge, $\rho_{s,p}$, is given as
\begin{eqnarray}
\rho_{s,p}=\frac{P}{1+\frac{\lambda_{1}}{t_{FE}\epsilon_{1}}+\frac{\lambda_{2}}{t_{FE}\epsilon_{2}}}, \label{eq7}
\end{eqnarray}
which is consistent with the common expression shown in Ref. \cite{PhysRevLett.94.246802}.

The energy band diagram is constructed by assuming that the bulk properties of metal electrodes remain the same under the applied bias; that is, the Fermi energy of the metal is fixed. Illustrated in Fig. \ref{fig3}(a) by setting the conduction band edge in the top metal contact as the zero energy reference, chemical potentials at top and bottom contacts ($\mu_{1}$ and $\mu_{2}$, respectively) have to satisfy the following equation:

\begin{eqnarray}
eV_{a} &=& \mu_{2} - \mu_{1} \nonumber \\
&=& \left(\frac{\rho_{s}\lambda_{1}}{\epsilon_{1}\epsilon_{0}} + \phi_{1} + E_{FE}t_{FE} - \phi_{2} - \frac{\rho_{s}\lambda_{2}}{\epsilon_{2}\epsilon_{0}} + E_{F2}\right)- E_{F1},  \nonumber \\
\label{eq20}
\end{eqnarray}
where $E_{F1}$ and $E_{F2}$ are Fermi energies of top and bottom metal electrodes, respectively.

\subsection{FTJ with Non-polar Dielectric}
\begin{figure}[h!]
\begin{center}
\subfloat[]{%
  \includegraphics[width = 1.9in]{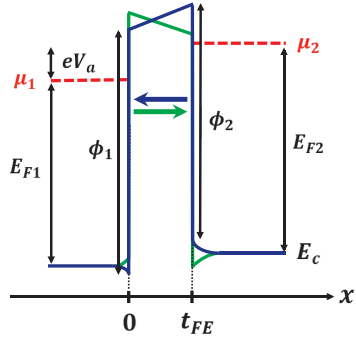}%
} \\
\subfloat[]{%
  \includegraphics[width = 1.9in]{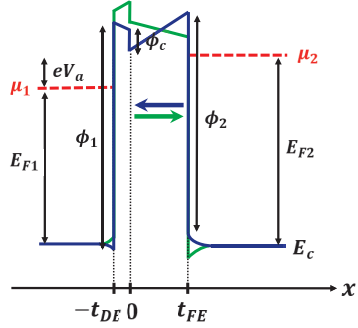}%
}
\caption{Schematics of energy band diagrams at a bias voltage $V_{a}$, satisfying $\mu_{2}-\mu_{1}=eV_{a}$, for FTJs (a) without and (b) with a non-polar DE layer between the FE and metal electrode. Arrows in the FE represent the direction of the electric polarization.}
\label{fig3}
\end{center}
\end{figure}

As a non-polar DE layer is presented between the top electrode and the FE as shown in Fig. \ref{fig1}(b), similar procedures to Section II. A can be followed to obtain the electric fields and potential profiles in an FTJ. Again by assuming that the electric displacement is continuous at interfaces, and the net voltage drop has to be entirely shared within the device, the following equations are satisfied.
\begin{eqnarray}
\rho_{s}&=&\epsilon_{0}E_{FE}+P=\epsilon_{0}\epsilon_{DE}E_{DE}, \label{eq8} \\
V_{a}+V_{bi}&=&\frac{\rho_{s}\lambda_{1}}{\epsilon_{1}\epsilon_{0}}+\frac{\rho_{s}\lambda_{2}}{\epsilon_{2}\epsilon_{0}}+E_{FE}t_{FE}+E_{DE}t_{DE}, \label{eq9} 
\end{eqnarray}
where $E_{DE}$ is the electric field across the DE, and $\epsilon_{DE}$ is the dielectric constant of the non-polar layer. By solving Eqs. \ref{eq8} and \ref{eq9}, the interface screening charge density and electric fields across the FE and the non-polar DE are given as
\begin{eqnarray}
\rho_{s} &=& \frac{\frac{\epsilon_{0}}{t_{FE}}\left(V_{a}+V_{bi}\right)+P}{1+\frac{t_{DE}}{\epsilon_{DE}t_{FE}}+\frac{\lambda_{1}}{\epsilon_{1}t_{FE}}+\frac{\lambda_{2}}{\epsilon_{2}t_{FE}}}, \label{eq10} \\
E_{FE}&=&\frac{\rho_{s}-P}{\epsilon_{0}}, \label{eq11} \\
E_{DE}&=&\frac{\rho_{s}}{\epsilon_{DE}\epsilon_{0}}, \label{eq12}
\end{eqnarray}
where $V_{bi}$ now is defined as $\frac{\left(\phi_{2}+\phi_{c}-\phi_{1}\right)}{e}$ with $\phi_{c}$ being the band discontinuity at the FE/non-polar DE interface. Note that the screening charge density induced solely by the FE bound charge can be obtained by removing both $V_{a}$ and $V_{bi}$ in Eq. \ref{eq10}, and the resulting expression is consistent with that in Ref. \cite{:/content/aip/journal/apl/95/5/10.1063/1.3195075}. After knowing the incomplete screening charge at the interface, the corresponding depolarization field can be calculated using Eq. \ref{eq11} and is given as
\begin{eqnarray}
E_{dep}=\frac{-P\left(\frac{t_{DE}}{\epsilon_{DE}}+\frac{\lambda_{1}}{\epsilon_{1}}+\frac{\lambda_{2}}{\epsilon_{2}}\right)}{\epsilon_{0}\left(t_{FE}+\frac{t_{DE}}{\epsilon_{DE}}+\frac{\lambda_{1}}{\epsilon_{1}}+\frac{\lambda_{2}}{\epsilon_{2}}\right)}.
\label{eq13}
\end{eqnarray}
As expected, Eq. \ref{eq13} is reduced to Eq. \ref{eq6} when $t_{DE}$ is reduced to zero. Similarly, by using the same energy reference in the previous case, the FTJ energy band diagram with a non-polar DE layer, as shown in Fig. \ref{fig3}(b), is established by satisfying the following equation:
\begin{eqnarray}
eV_{a} &=& \mu_{2} - \mu_{1} \nonumber \\
&=& \left(\frac{\rho_{s}\lambda_{1}}{\epsilon_{1}\epsilon_{0}} + \phi_{1} + E_{DE}t_{DE} - \phi_{c} + E_{FE}t_{FE}\right. \nonumber \\
& & \left. - \phi_{2} - \frac{\rho_{s}\lambda_{2}}{\epsilon_{2}\epsilon_{0}} + E_{F2}\right)- E_{F1}. \label{eq21}
\end{eqnarray}

\subsection{FE Hysteresis Loop}
To describe the electric polarization response of a FE thin film under applied bias, built-in field, and depolarization field, the Landau-Khalatnikov (LK) equation is used and given as \cite{PSSB:PSSB765}
\begin{eqnarray}
\gamma\frac{\partial P}{\partial t} = -\frac{\partial F}{\partial P} \label{eq14},
\end{eqnarray}
where $\gamma$ is the viscosity coefficient and $F$ is the FE free energy including the bulk and interactions with different types of electric fields, which can be in general expanded in terms of the thermodynamic order parameter based on the Landau theory and is written as
\begin{eqnarray}
F &=& \alpha_{1}P^{2} + \alpha_{11}P^{4} + \alpha_{111}P^{6} - \frac{1}{2}E_{dep}P \nonumber \\
&-& \left(E_{FE}-E_{dep}\right)P
\label{eq15}
\end{eqnarray}
with $\alpha_{1}$, $\alpha_{11}$, and $\alpha_{111}$ being free energy expansion coefficients \cite{PhysRevB.68.094113,PhysRevApplied.4.044014,7373582,PhysRevB.88.024106}. The contribution from both built-in and applied electric fields is included in the last term in Eq. \ref{eq15}. 

While Ref. \cite{PhysRevB.68.094113} pointed out that Eq. \ref{eq14} is particularly for the intrinsic single-domain FE switching, which typically requires a defect-free FE thin film with a very small cross-sectional area and is quite different from the extrinsic switching driven by FE domain nucleation and propagation, here for simplicity, we assume that the electric polarization in a FE thin film can be represented by an effective electric polarization, $P$, satisfying the LK equation, and the experimental FE hysteresis loops, characterized by the remanent polarization and coercive voltage, can be well described by adjusting expansion and viscosity coefficients. Furthermore, by using Eq. \ref{eq14}, the shift in a FE hysteresis loop due to a non-zero built-in field across a FTJ can also be easily captured \cite{PhysRevB.88.024106}. Note that typically the electric displacement through the FE, $D$, is written as \cite{:/content/aip/journal/jap/72/12/10.1063/1.351910}
\begin{eqnarray}
D=\epsilon_{0}\left(1+\chi\right)E_{FE}+P_{d}, \label{eq16}
\end{eqnarray}
where $\chi$ accounts for the linear contribution of the polarization and $P_{d}$ is the polarization due to switching dipoles. However, in the LK equation mentioned above, $P$ accounts for the effects from both linear response and switching dipoles, and thus the electric displacement is simply written as $\epsilon_{0}E_{FE}+P$.

\subsection{Tunneling Currents}
As shown in Fig. \ref{fig3}, based on Eqs. \ref{eq20} and \ref{eq21}, the energy band diagram can be constructed for a given electric polarization obtained from the LK equation and is used as the electron potential energy in the non-equilibrium Green's function (NEGF) method to calculate the transmission coefficient \cite{Supriyo_2005}. For the tunneling currents, the Landau formula is applied and given as \cite{5392683}
\begin{eqnarray}
J = -\sum_{k_{y},k_{z}}\frac{2e}{Ah}\int dE t(E)\left\lbrace f_{1}\left(E\right) - f_{2}\left(E\right)\right\rbrace, \label{eq17}
\end{eqnarray}
where $k_{y}$ and $k_{z}$ are electron wave vectors in the transverse plane, $e$ is the elementary charge, $A$ is the cross-sectional area, $E$ is the total electron energy, $t$ is the transmission coefficient, and $f_{1}$ and $f_{2}$ are Fermi-Dirac distributions for top and bottom metal contacts, respectively, given as
\begin{eqnarray}
f_{1\left(2\right)}\left(E\right)=\frac{1}{1+e^{\frac{E-\mu_{1\left(2\right)}}{k_{B}T}}}, \label{eq18}
\end{eqnarray}
where $\mu_{1}$ and $\mu_{2}$ are chemical potentials of top and bottom metal contacts with $\mu_{2}-\mu_{1}=eV_{a}$, $k_{B}$ is the Boltzmann constant, and $T$ is the temperature. The details of writing an alternative expression for currents using the electron wave vector in the spherical coordinate are shown in Appendix B. The transmission coefficient in Eq. \ref{eq17} is calculated using the Green's function, $\bf{G}$, given as
\begin{eqnarray}
t=trace\left(\bf{\Gamma^{t}G\Gamma^{b}G^{\dagger}}\right), \label{eq19}
\end{eqnarray}
where $\bf{G}$ is defined as $\bf{\left(EI-H-\Sigma_{t}-\Sigma_{b}\right)}^{-1}$ with $\bf{I}$, $\bf{H}$, and $\bf{\Sigma}$ being the identity matrix, device Hamiltonian, and contact self-energy, respectively, and $\bf{\Gamma}$ is the broadening function defined as $i\left(\bf{\Sigma-\Sigma^{\dagger}}\right)$. The detailed expression of the Hamiltonian and contact self-energy can be found in the Appendix C.

\section{Results and Discussion}
In this section, the theoretical framework presented above is used to explain existing experimental results \cite{Chanthbouala2012,doi:10.1021/nl302912t,Tian2016}. First, to show the model captures key underlying physics behind FTJs, measured $I$-$V$ characteristics for both inorganic and organic FTJs are fitted by using proper energy band diagram and LK parameters. Next, the concept of effective screening length and dielectric constant is applied to explain the opposite high/low resistance states observed in Co/BTO/LSMO systems \cite{Chanthbouala2012,doi:10.1021/nl302912t}, which may result from interface termination effects \cite{ADFM:ADFM201500371}. Finally, the model is extended by including a CoO$_{x}$ non-polar buffer layer at the Co/BTO interface, and it is shown that the voltage-dependent oxygen vacancies at the CoO$_{x}$/BTO interface may be partially responsible for the memristor behavior as mentioned in Ref. \cite{doi:10.1021/nl302912t}.

\subsection{Comparison with Experimental $I$-$V$ Characteristics}
In this paper, for an FTJ, it is assumed that TER is a main consequence of modifying the energy band diagram through depolarization fields induced by incomplete screening charge at FE/metal interfaces, and is expected to vary with the polarization. In other words, at a given voltage, a larger difference in two opposite polarization states leads to more pronounced TER. Hence, to describe measured FTJ $I$-$V$ characteristics, it is required to accurately model FE hysteresis loops, which are also presented in the following comparisons with experiments. Note that, for simplicity, all the FE hysteresis loops in this work are simulated by applying a sinusoidal voltage signal with a period of $70$ ps, and LK parameters are adjusted accordingly to obtain a reasonable FE response observed in experiments. In reality, FE thin films may have different dynamic responses with respect to an applied bias, depending on the quality, material, or size of the sample. 

\begin{figure}[h!]
\begin{center}
\subfloat[]{%
  \includegraphics[width = 3.2in]{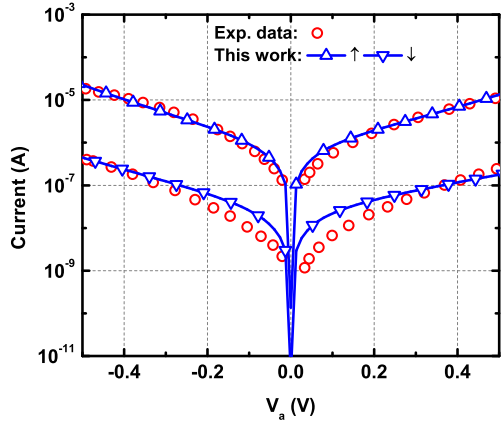}%
} \\
\subfloat[]{%
  \includegraphics[width = 3.2in]{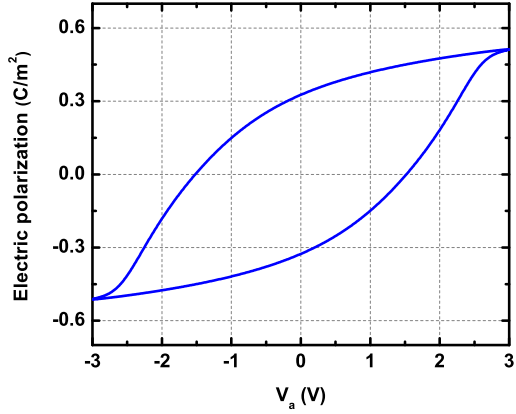}%
}
\caption{(a) Comparison between FTJ (Co/BTO/LSMO) experimental data \cite{Chanthbouala2012} and simulation results using the following band diagram parameters: $t_{FE}=2$ nm, $\phi_{1}=\phi_{2}=7.15$ eV, $E_{F1}=E_{F2}=6.5$ eV, $\epsilon_{1}=2.5$, $\epsilon_{2}=9.8$, $\lambda_{1}=0.5\times 10^{-10}$ m \cite{0022-3727-47-4-045001}, $\lambda_{2}=1\times 10^{-10}$ m \cite{0022-3727-47-4-045001}, $m^{*}=0.8m_{0}$. (b) Simulated FE hysteresis loop for FTJ (Co/BTO/LSMO) experiments \cite{Chanthbouala2012} ($V_{c}\sim\pm 3$ V, $\epsilon_{FE}\sim 15$, and $P_{r}\sim 0.3$ C/m$^{2}$) with the following LK parameters: $\gamma=10^{-2}$ m sec/F, $\alpha_{1}=-2.77\times 10^{7}$ m/F, $\alpha_{11}=-5.35\times 10^{8}$ m$^{5}/$C$^{2}$F, and $\alpha_{111}=6.4\times 10^{9}$ m$^{9}/$C$^{4}$F.}
\label{fig4}
\end{center}
\end{figure}

First, the measured FTJ $I$-$V$ characteristics in a Co/BTO/LSMO layered structure \cite{Chanthbouala2012} are used to justify our theoretical approach. Since there is no clear shift in hysteresis loops observed in experiments, it is assumed that a built-in field across the junction is close to zero, which implies $\phi_{1}$ is equal to $\phi_{2}$ in our model. Next, LK parameters for BTO \cite{PhysRevApplied.4.044014} are slightly varied so that the FE thin film exhibits a hysteresis loop with $V_{c}\sim\pm 3$ V, $\epsilon_{FE}\sim 15$, and $P_{r}\sim 0.3$ C/m$^{2}$ as shown in Fig. \ref{fig4}(b), where $V_{c}$, $\epsilon_{FE}$, and $P_{r}$ are the coercive voltage, the FE dielectric constant, and the remanent polarization, respectively. By assuming the following interface parameters: $\lambda_{1}=0.5\times 10^{-10}$ m \cite{0022-3727-47-4-045001}, and $\lambda_{2}=1\times 10^{-10}$ m \cite{0022-3727-47-4-045001}, $\phi_{1}$, $\phi_{2}$, $\epsilon_{1}$, $\epsilon_{2}$, and $m^{*}$ are varied to obtain a good agreement with experimental data as shown in Fig. \ref{fig4}(a), which shows that in Co/BTO/LSMO systems, a depolarization field modifying the energy band diagram is the dominant driving force for TER, rather than the effects due to strain \cite{PhysRevB.72.125341} or FE polarization dependent complex band structure \cite{PhysRevLett.98.137201}. However, even though the experimental data can be well described by depolarization fields in Fig. \ref{fig4}, changes in FTJ energy band diagrams through polarization reversals is not a pure charge-mediated (or electrostatic) effect.  This is mainly because the effective screening length and the dielectric response significantly depend on the specific interface geometry, which is a fully quantum-mechanical outcome and requires approaches in the microscopic level such as first-principles calculations \cite{Stengel2009}.

\begin{figure}[h!]
\begin{center}
\subfloat[]{%
  \includegraphics[width = 3.2in]{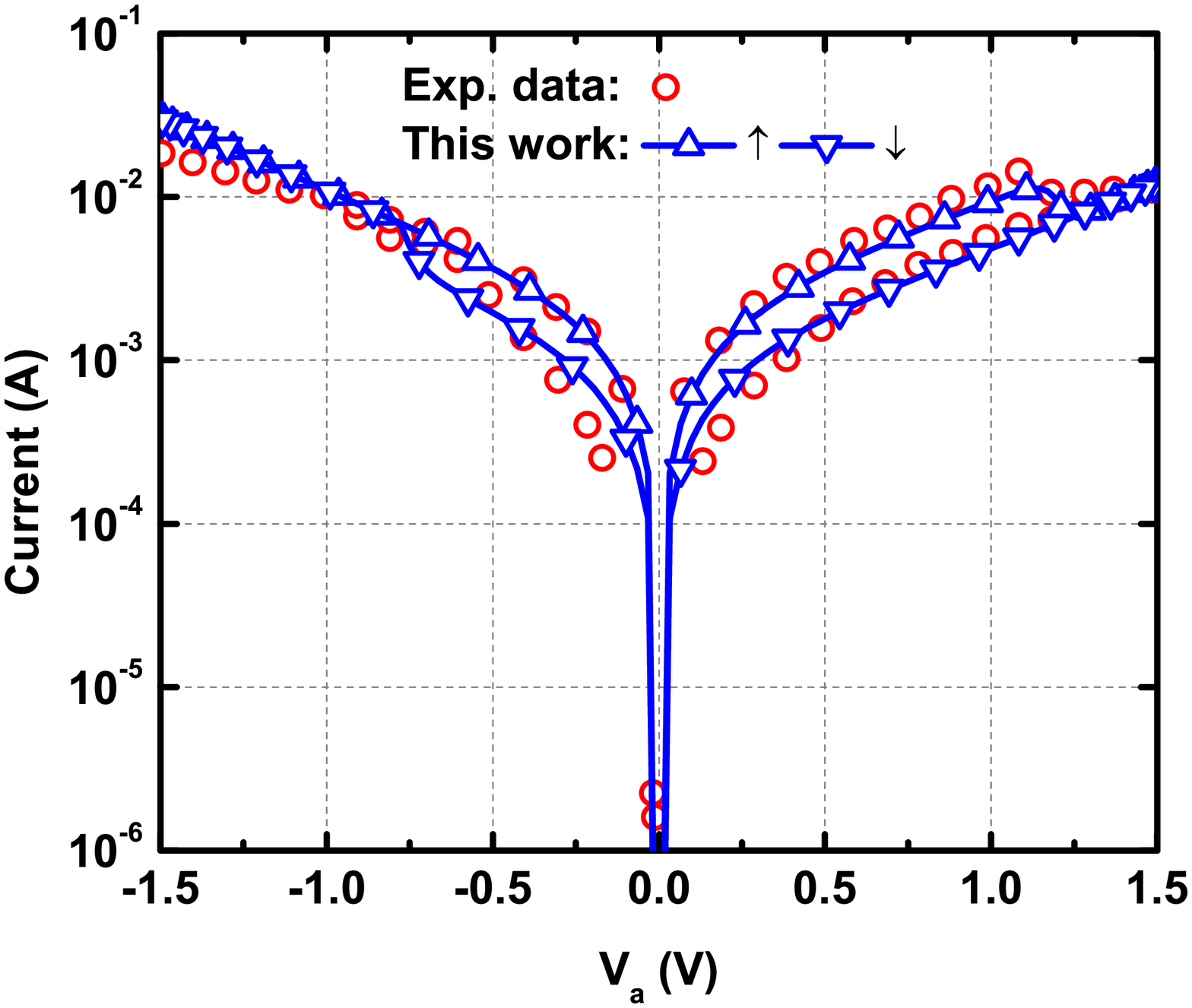}%
} \\
\subfloat[]{%
  \includegraphics[width = 3.2in]{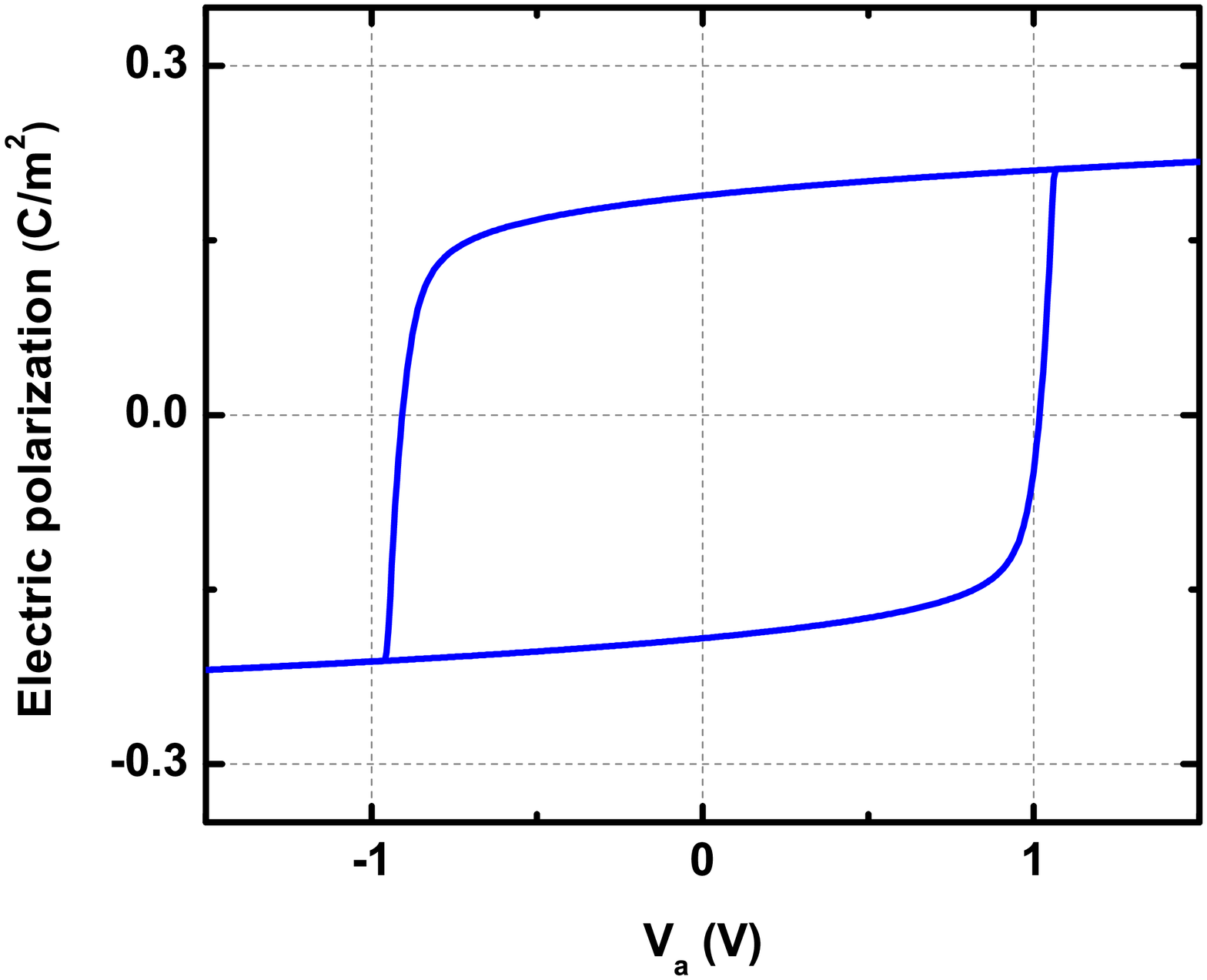}%
}
\caption{(a) Comparison between FTJ (Au/PVDF/W) experimental data \cite{Tian2016} and simulation results using the following band diagram parameters: $t_{FE}=2$ nm, $\phi_{1}=6.76$ eV, $\phi_{2}=6.7$ eV, $E_{F1}=E_{F2}=6.5$ eV, $\epsilon_{1}=6.5$, $\epsilon_{2}=20$, $\lambda_{1}=0.75\times 10^{-10}$ m \cite{Gajek2007}, $\lambda_{2}=0.45\times 10^{-10}$ m \cite{w_screening_length}, $m^{*}=0.1m_{0}$. (b) Simulated FE hysteresis loop for FTJ (Au/PVDF/W) experiments \cite{Tian2016} ($V_{c}\sim\pm 1$ V, $\epsilon_{FE}\sim 4.4$, and $P_{r}\sim 0.18$ C/m$^{2}$) with the following LK parameters: $\gamma=1.5\times 10^{-3}$ m sec/F, $\alpha_{1}=-1.38\times 10^{9}$ m/F, $\alpha_{11}=-2.67\times 10^{10}$ m$^{5}/$C$^{2}$F, and $\alpha_{111}=8\times 10^{11}$ m$^{9}/$C$^{4}$F.}
\label{fig5}
\end{center}
\end{figure}

In Fig. \ref{fig4}, since only the currents at low voltages are measured, the full dependence of tunneling currents on an FE hysteresis loop cannot be observed. As a result, an $I$-$V$ characteristic curve reported in an Au/poly-vinylidene fluoride (PVDF)/W layered structure is used to justify our model for a complete FE sweep \cite{Tian2016}. Again, to fit experimental data, a FE hysteresis loop of a monolayer PVDF film is generated by tuning LK parameters as shown in Fig. \ref{fig5}(b), in which the resulting $V_{c}$, $\epsilon_{FE}$, and $P_{r}$ are about $1$ V, $4.4$, and $0.18$ C/m$^{2}$, respectively. By using the following interface parameters: $\epsilon_{1}=6.5$, $\lambda_{1}=0.75\times 10^{-10}$ m \cite{Gajek2007}, and $\lambda_{2}=0.45\times 10^{-10}$ \cite{w_screening_length}, $\phi_{1}$, $\phi_{2}$, $\epsilon_{2}$, and $m^{*}$ are adjusted to match experimental data as shown in Fig. \ref{fig5}(a), where a good agreement between the theoretical and experimental results is reached. Note that a weak built-in field, observed in the experiment \cite{Tian2016} and leading to a small shift in the hysteresis loop as shown in Fig. \ref{fig5}(b), is included to obtain a better fit to the experimental data.

In Fig. \ref{fig5}(a), it can be seen that TER varies laregly with the electric polarization; that is, the difference between high and low resistance states is reduced as the voltage is close to or beyond the coercive voltage. Furthermore, since the interface parameters for Fig. \ref{fig5}(a) are more close to bulk values, it can also be concluded that TER in an Au/PVDF/W organic FTJ is more dominated by a pure electrostatic effect, rather than complex changes of interfacial bonds, which can be attributed to the fact that the electrodes are attached to PVDF thin films using mainly Van der Waals forces in an Au/PVDF/W structure \cite{Tian2016}.

\subsection{Interface Termination Effects on TER}
\begin{figure}[h!]
\begin{center}
\subfloat[]{%
  \includegraphics[width = 3.2in]{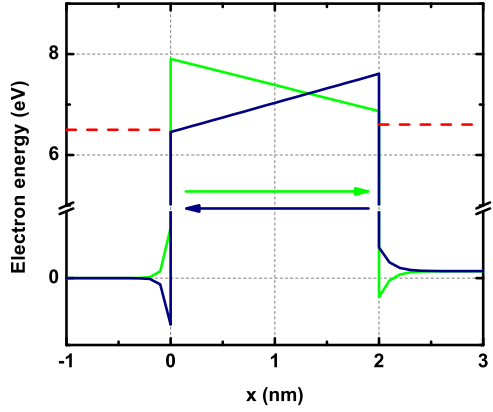}%
} \\
\subfloat[]{%
  \includegraphics[width = 3.2in]{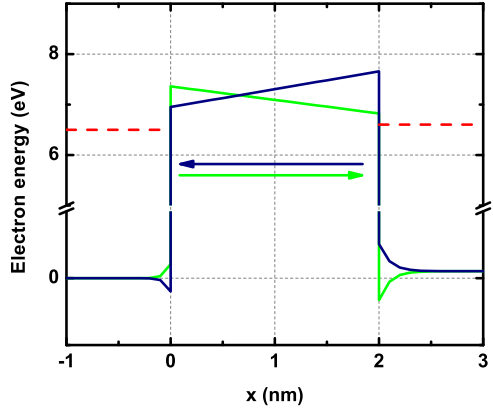}%
}
\caption{Energy band diagrams at $0.1$ V for both polarization states with two different effective contact ratios: (a) $1.98$ and (b) $0.49$. The dark blue and the green correspond to the polarization states pointing to the top and bottom contacts, respectively. Red dash lines represent chemical potentials at both contacts.}
\label{fig7}
\end{center}
\end{figure}

As shown in Figs. \ref{fig4} and \ref{fig5}, in both experiments \cite{Chanthbouala2012,Tian2016}, the low and high resistance states correspond to the electric polarizations pointing to the top (Co or Au) and the bottom (LSMO or W) electrodes, respectively. These experimental results can be explained by the energy band diagram shown Fig. \ref{fig7}(a), where a lower tunnel barrier is produced as the polarization is pointing to the top contact, which has larger changes in the interface potential energy. Note that as shown in Eqs. \ref{eq1} and \ref{eq2}, a higher ratio of $\frac{\lambda}{\epsilon}$ leads to a larger change in the interface potential energy. From Fig. \ref{fig7}(a), it is found that since at low voltages, the energy slope on the FE barrier is mostly dominated by the depolarization field, whose direction is always opposite to that of the polarization, the top and bottom interfaces have opposite effects on the tunnel barrier. Using the polarization pointing to the top contact as an example, the top and bottom interface potential changes reduce and increase the FE barrier, respectively, and these contact effects on the barrier are reversed as the polarization is switched to the opposite direction. Consequently, if the interface energy change at the top is greater than that at the bottom, the FE barrier for the polarization pointing to the top will be lower and thus a lower resistance state is generated. Therefore, as shown in Fig. \ref{fig7}(a), it seems that interface quantities play a significant role in determining the relation between the high/low resistance states and the polarization direction. Here a quantity called the effective contact ratio is defined as $\frac{\lambda_{1}\epsilon_{2}}{\lambda_{2}\epsilon_{1}}$ to distinguish the high/low resistance states in an FTJ. In Figs. \ref{fig4} and \ref{fig5}, the effective contact ratios are $1.96$ and $5.1$, respectively, which are both larger than $1$, implying that the resistance states are more dominated by the top interface. As a result, the lower resistance state is for the polarization pointing to the top contact (or the TER sign is "$+$"), consistent with experimental observations. 

In Au/PVDF/W FTJs, it is believed that a depolarization field creates larger changes in the potential energy at the Au side \cite{Tian2016}, and so far, no experimental evidence has shown that high/low resistance states can be switched in the same FTJ structure, which is probably because contacts and an organic FE film are attached through Van der Waals forces, rather than complex interface bonds as mentioned previously \cite{Tian2016}. However, in Co/BTO/LSMO layered structures, several groups have reported an opposite relation between the polarization direction and the resistance state \cite{Chanthbouala2012,doi:10.1021/nl302912t}. Recently, some groups have reported that the reversal of the high/low resistance states in Co/BTO/LSMO systems is attributed to either TiO$_{2}$ or BaO terminated at the Co/BTO interface \cite{ADFM:ADFM201500371}. To support this argument theoretically, our model provides an intuitive picture for the reversal of high/low resistance states induced by termination effects. As predicted by first-principles calculations, the screening length is almost zero at the Co/TiO$_{2}$-terminated BTO interface \cite{Stengel2009}. Therefore, in Fig. \ref{fig7}(b), the effective contact ratio is set to be less than $1$ without adjusting $\frac{\lambda}{\epsilon}$ of the bottom interface, and it is shown that compared to Fig. \ref{fig7}(a), where the effective contact ratio is larger than $1$, a lower tunneling barrier is generated by the polarization pointing the bottom electrode, rather than the top one, and thus the high/low resistance states are reversed.

\begin{figure}[h!]
\begin{center}
\subfloat[]{%
  \includegraphics[width = 3.2in]{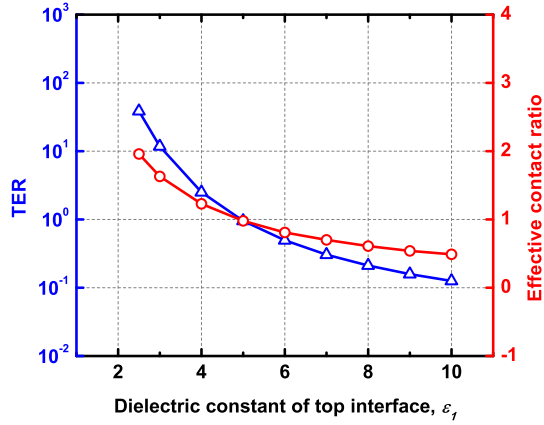}%
} \\
\subfloat[]{%
  \includegraphics[width = 3.2in]{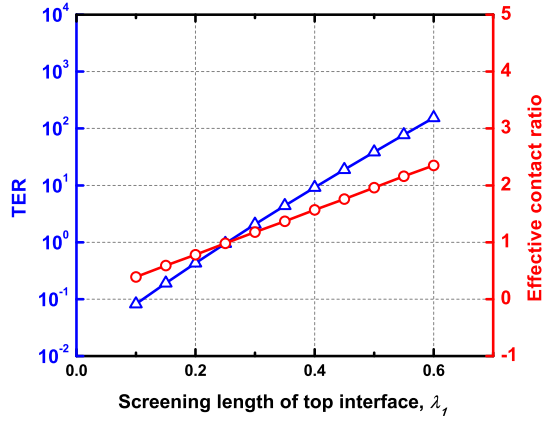}%
}
\caption{TER at $V_{a} = 0.1$ V and effective contact ratio versus top contact (a) dielectric constant and (b) screening length. TER and effective contact ratios are defined as $\frac{I_{\uparrow}}{I_{\downarrow}}$ and $\frac{\lambda_{1}\epsilon_{2}}{\lambda_{2}\epsilon_{1}}$, respectively.}
\label{fig6}
\end{center}
\end{figure}

Figs. \ref{fig6}(a) and (b) clearly indicate that rather than the individual interface properties, the effective contact ratio is the most essential factor to determine both sign and magnitude of TER, defined as $\frac{I_{\uparrow}}{I_{\downarrow}}$, where $I_{\uparrow}$ and $I_{\downarrow}$ are the currents corresponding to the polarizations pointing to the top and bottom electrodes, respectively. In Figs. \ref{fig6}(a) and (b), it is shown that a more pronounced TER can be produced as the top and the bottom interfaces become more distinct ($\frac{\lambda_{1}\epsilon_{2}}{\lambda_{2}\epsilon_{1}}\gg 1$ or $\ll 1$). Also, from the same figures, a lower resistance state is always produced by the polarization pointing to the interface with larger $\frac{\lambda}{\epsilon}$ as explained in Figs. \ref{fig7}(a) and (b). In other words, the sign of TER, as it is defined here, is switched from "$+$" to "$-$" as the effective contact ratio changes from the value larger than $1$ to less than $1$. As a result, if the effective contact ratio is equal to $1$, meaning that the device is perfectly symmetric, the resulting TER will also be $1$, and thus it is impossible to distinguish the polarization direction through tunneling resistance.

\subsection{FTJs with CoO$_{x}$}
From the previous section, it is shown that TER significantly depends on metal/BTO interface properties in an FTJ. Moreover, in addition to the termination effect, recently some experimental studies have reported that an inevitable CoO$_{x}$ layer at the Co/BTO interface plays an important role for the memristor behavior of a Co/BTO/LSMO FTJ; that is, TER varies with the magnitude of the writing voltage \cite{doi:10.1021/nl302912t}. Hence, in this section, our simple model is extended as shown in Figs. \ref{fig1}(b) and \ref{fig3}(b) to investigate the CoO$_{x}$ effect on TER.

\begin{figure}[h!]
\begin{center}
\subfloat[]{%
  \includegraphics[width = 3.2in]{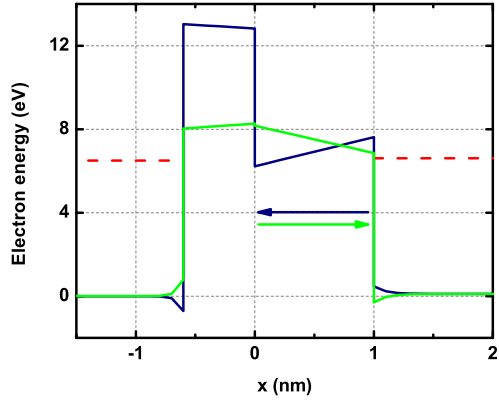}%
} \\
\subfloat[]{%
  \includegraphics[width = 3.2in]{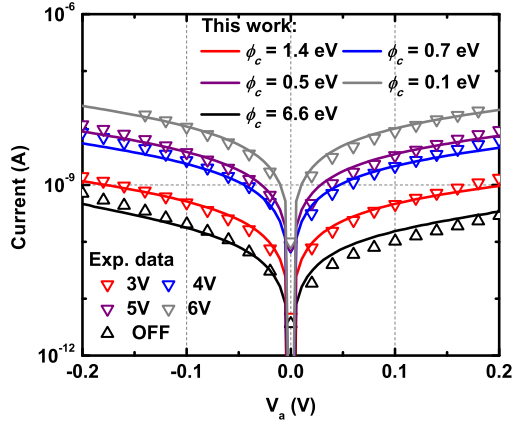}%
}
\caption{(a) Energy band diagrams at $0.1$ V for high/low resistance states in an FTJ with a CoO$_{x}$ buffer layer at the interface. $\phi_{c}$ for high and low resistance states are $6.6$ and $0.1$ eV, respectively. (b) Comparison with experimental data \cite{doi:10.1021/nl302912t} using various $\phi_{c}$ for high and low resistance states and different writing voltages. In addition to $t_{DE} = 0.6$ nm, $t_{FE} = 1$nm, $\phi_{1}$ and $\phi_{c}$, the simulation parameters are the same as those in Fig. \ref{fig4}.}
\label{fig9}
\end{center}
\end{figure}

As mentioned in Ref. \cite{doi:10.1021/nl302912t}, a positive (negative) applied bias accumulates (dissipates) oxygen vacancies at the CoO$_{x}$/BTO interface, effectively reducing (increasing) $\phi_{c}$. Therefore, as shown in the energy band diagrams of Fig. \ref{fig9}(a), which are constructed using Eq. \ref{eq21}, the low (high) resistance state corresponds to the polarization pointing to the bottom (top) contact with smaller (larger) $\phi_{c}$. Note that as predicted in Ref. \cite{:/content/aip/journal/apl/95/5/10.1063/1.3195075}, an unchanged $\phi_{c}$ in both polarization directions will result in a reversal of high/low resistance states, which haven't been observed in the experiment yet \cite{doi:10.1021/nl302912t}. Furthermore, since no significant shift in the FE hysteresis loop was observed in the experiment \cite{doi:10.1021/nl302912t}, in our model, $\phi_{1}$ is adjusted accordingly with $\phi_{c}$ so that the built-in field across the device is zero. In other words, $\phi_{2}+\phi_{c}-\phi_{1}=0$, where $\phi_{2}$ is fixed due to no change at the BTO/LSMO interface. Therefore, by using the same simulation parameters for the interfaces and the FE hysteresis loop as listed in Fig. \ref{fig4}, and assuming that part of BTO transforms into CoO$_{x}$ ($t_{DE} = 0.6$ nm and $t_{FE} = 1$ nm), $\phi_{c}$ is adjusted to fit the experimental data as shown in Fig. \ref{fig9}(b), where a good agreement between the theory and the experiment is reached. As a result, Fig. \ref{fig9}(b) shows that it is possible to change TER through modifications of $\phi_{c}$ induced by voltage-dependent oxygen vacancies at the CoO$_{x}$/BTO interface. However, it seems that the required change in $\phi_{c}$ from off to on states may be too drastic for simply the charge-mediated effect ($6.6$ to $0.1$eV). Therefore, the thickness of CoO$_{x}$ may also be altered depending on the applied bias; that is, the CoO$_{x}$ thickness may be reduced (increased) as the FTJ is switched from high (low) to low (high) resistance states. More experimental studies are required to confirm the possibility of the voltage-dependent CoO$_{x}$ thickness in an FTJ.

\section{Conclusion}
This paper presents a theoretical description of quantum-mechanical electronic transport and thermodynamic ferroelectric responses in both organic and inorganic FTJs. Inversed TER effect with respect to the polarization direction reported by different groups can also be explained by the proposed model through the effective contact ratio and termination effects. Finally, the role of a CoO$_{x}$ buffer layer at the Co/BTO interface is also examined. It is found that the sizable memristive effects cannot be explained solely by the change in the barrier height due to charge-mediated effects. It is suggested that the CoO$_{x}$ layer thickness may also change as a result of electrically-induced Co oxidation/reduction at the Co/BTO interface. The proposed approach for description of the electroresistance effect in FTJs will provide a foundation for performance optimization of the core elements for nonvolatile memory and logic devices.

\begin{acknowledgements}
This work is sponsored by Semiconductor Research Corporation NRI Theme 2624.001 and 2398.002. A. Gruverman also acknowledges support by the National Science Foundation (NSF) under Grant ECCS-1509874.
\end{acknowledgements}

\begin{figure}[h!]
\begin{center}
\includegraphics[width=2.5in]
{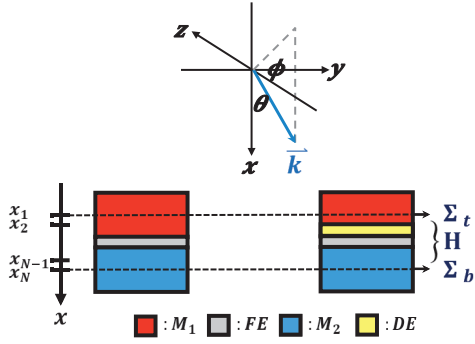}
\caption{Schematics of illustrating the electron wave vector in the spherical coordinate and the non-equilibrium Green's function (NEGF) approach to FTJs without and with a non-polar DE layer between the FE and metal electrode.}
\label{figA}
\end{center}
\end{figure}

\appendix
\section{Derivation of Eqs. \ref{eq2} and \ref{eq3}}
The relation between charge ($Q$) and electric field ($E$) in the metal can be described by the Poisson's equation given as
\begin{eqnarray}
\frac{\partial E\left(x\right)}{\partial x}=\frac{Q}{\epsilon_{m}\epsilon_{0}}=\frac{-e\left( n-n_{0}\right)}{\epsilon_{m}\epsilon_{0}},
\label{eqA1}
\end{eqnarray}
where $\epsilon_{m}$ is the dielectric constant of the metal, $n$ is the electron density, and $n_{0}$ is the electron density in the neutral metal electrode. In the metal, the electrons can be treated as a free fermi gas, and thus the local potential ($V$) and electron density can be related as \cite{Kittel:ISSP}
\begin{eqnarray}
V=\frac{\hbar^{2}}{2m_{0}}\left(3\pi^{2}n\right)^{\frac{2}{3}}
\end{eqnarray}
with $\hbar$ being the reduced Planck constant, and $m_{0}$ being the free electron mass. By using $-\frac{\partial V}{\partial x}=E$, the derivative of the electron density with respect to $x$ can be expressed as
\begin{eqnarray}
\frac{\partial n}{\partial x}=-\frac{E}{\frac{\hbar^{2}}{3m_{0}}\left(3\pi^{2}\right)^{\frac{2}{3}}n^{\frac{-1}{3}}},
\end{eqnarray}
and therefore the derivative of Eq. \ref{eqA1} with respect to $x$ becomes
\begin{eqnarray}
\frac{\partial^{2} E}{\partial x^{2}}&=&\frac{-e}{\epsilon_{m}\epsilon_{0}}\frac{\partial n}{\partial x}=\frac{E}{\lambda^{2}},
\label{eqA2}
\end{eqnarray}
where the metal Thomas-Fermi screening length, $\lambda$, is defined as $\frac{\hbar^{2}\epsilon_{m}\epsilon_{0}}{3em_{0}}\left(3\pi^{2}\right)^{\frac{2}{3}}n^{\frac{-1}{3}}$. The general solution of Eq. \ref{eqA2} is $Ae^{\frac{x}{\lambda}}+Be^{\frac{-x}{\lambda}}$ with $A$ and $B$ being coefficients determined by the boundary conditions, which are, using the top electrode as an example, $E\left(-\infty\right)=0$ and $E\left(0\right)=\frac{\rho_{s}}{\epsilon_{1}\epsilon_{0}}$. Therefore, the corresponding electric field ($E_{1}$) and potential profile ($V_{1}$) ($-\infty < x \leq 0$) are given as
\begin{eqnarray}
E_{1}&=&\frac{\rho_{s}}{\epsilon_{1}\epsilon_{0}}e^{\frac{x}{\lambda_{1}}}, \\
V_{1}&=&-\int_{-\infty}^{x}dx'E\left(x'\right)=\frac{-\rho_{s}\lambda_{1}}{\epsilon_{1}\epsilon_{0}}e^{\frac{x}{\lambda_{1}}}.
\label{eqA3}
\end{eqnarray}
Similarly, by using $E\left(\infty\right)=0$ and $E\left(t_{FE}\right)=\frac{-\rho_{s}}{\epsilon_{2}\epsilon_{0}}$ as boundary conditions, the potential profile ($V_{2}$) of the bottom electrode ($t_{FE} \leq x < \infty$) is given as
\begin{eqnarray}
V_{2}=\frac{\rho_{s}\lambda_{2}}{\epsilon_{2}\epsilon_{0}}e^{\frac{-\left(x-t_{FE}\right)}{\lambda_{2}}}.
\label{eqA4}
\end{eqnarray}
Eqs. \ref{eqA3} and \ref{eqA4} are identical to Eqs. \ref{eq1} and \ref{eq2}. Note that as mentioned in the main text, for some FTJs with complex interfacial bonds, the potential drop near the interface is described by the effective screening length and dielectric response, rather than the Thomas-Fermi one \cite{Junquera:2008-11-01T00:00:00:1546-1955:2071}.

\section{Alternative Expression of Eq. \ref{eq16}}
The electron wave vector can be represented in the spherical coordinate as shown in Fig. \ref{figA}. To rewrite Eq. \ref{eq16}, the first step is to convert the summation into the integral using periodic boundary conditions ($\sum_{k}=\frac{L}{2\pi}\int dk$), and the resulting expression is given as
\begin{eqnarray}
J=\frac{-e}{2\pi^{2}h}\int_{-\infty}^{\infty}\int_{-\infty}^{\infty}d k_{y} d k_{z}\int dE t\left( f_{1}-f_{2}\right).
\end{eqnarray}
Note that $t$, $f_{1}$, and $f_{2}$ are all energy-dependent. Under the spherical coordinate, $dk_{y}dk_{z}$ can be written as $k^{2}\sin\theta d\phi d\theta$. For electrons coming from $+x$ with total energy, $E$, equal to $E=\frac{\hbar^{2}k^{2}}{2m^{*}}+U_{0}$, where $m^{*}$ is the effective mass and $U_{0}$ is the potential energy, the current equation becomes
\begin{eqnarray}
J&=&\frac{-e}{2\pi^{2}h}\int_{0}^{2\pi}\int_{0}^{\frac{\pi}{2}} d\phi d\theta k^{2}\sin\theta \int dE t\left( f_{1}-f_{2}\right) \nonumber \\
&=&\frac{-em}{\pi^{2}\hbar^{3}}\int_{0}^{\frac{\pi}{2}}d\theta \sin\theta \int_{U_{0}}^{\infty} dE \left(E-U_{0}\right)t\left(f_{1}-f_{2}\right). \nonumber \\
\label{eqB1}
\end{eqnarray}
It can be seen from Eq. \ref{eqB1} that the tunneling currents account for all the contribution of electrons from different energy levels and injection angles in the metal contact. 

\section{Device Hamiltonian and Contact Self-energy}
The device Hamiltonian, $\bf{H}$, is constructed based on a single-band effective mass Hamiltonian operator of an electron given as
\begin{eqnarray}
\hat{H}=\frac{-\hbar^{2}}{2m^{*}}\frac{\partial^{2}}{\partial x^{2}}+\frac{\hbar^{2}\left(k_{y}^{2}+k_{z}^{2}\right)}{2m^{*}} + U\left(x\right), \\
\nonumber
\end{eqnarray}
where $U\left(x\right)$ is the energy band diagram of an FTJ. Note that in this approach, a space-independent effective mass, $m^{*}$, is used to characterize the quantum-mechanical tunneling process in the thin-film device. Therefore, by considering an electron coming from $+x$ with total energy, $E$, equal to $E=\frac{\hbar^{2}k^{2}}{2m^{*}}+U_{0}$, the operator can be rewritten using Fig. \ref{figA} and is given as
\begin{eqnarray}
\hat{H}&=&\frac{-\hbar^{2}}{2m^{*}}\frac{\partial^{2}}{\partial x^{2}}+ \left(E-U_{0}\right)\sin^{2}\theta + U\left(x\right) \nonumber \\
&=&\frac{-\hbar^{2}}{2m^{*}}\frac{\partial^{2}}{\partial x^{2}}+ E_{\perp}\left(\theta\right) + U\left(x\right),
\end{eqnarray}
where $E_{\perp}$ is the transverse energy of the electron, which depends on the injection angle, $\theta$. The device Hamiltonian can be obtained by simply converting $\hat{H}$ into a matrix using the finite difference method and is given as
\begin{widetext}
\begin{eqnarray}
\bf{H}=\left[
\begin{array}{c c c c c}
2t+E_{\perp}\left(\theta\right)+U\left(x_{1}\right) & -t & 0 & \cdots & \cdots \\
-t & 2t+E_{\perp}\left(\theta\right)+U\left(x_{2}\right) & -t & 0 & \cdots \\
\vdots & \vdots & \ddots & \ddots & \ddots \\
0 & 0 & \cdots & \cdots & 0 \\
0 & 0 & 0 & \cdots & \cdots
\end{array}
\right. \\
\left.
\begin{array}{c c c}
0 & 0 & 0 \\
\cdots & 0 & 0 \\
\ddots & \vdots & \vdots \\
-t & 2t+E_{\perp}\left(\theta\right)+U\left(x_{N-1}\right) & -t \\
0 & -t & 2t+E_{\perp}\left(\theta\right)+U\left(x_{N}\right)
\end{array}
\right], \nonumber
\end{eqnarray}
\end{widetext}
where the $x$ axis is divided into $N$ mesh points, $x_{1}$, $x_{2}$, $\cdots$, $x_{N-1}$, and $x_{N}$, and $t$ is the coupling strength between the nearest neighbors defined as $t=\frac{\hbar^{2}}{2m^{*}a^{2}}$ with $a$ being the distance between two nearest mesh points, which is set as $0.1$ nm in the main text. Under the open boundary condition, the self-energies of top and bottom contacts are given as
\begin{eqnarray}
\bf{\Sigma_{t}}=\left[
\begin{array}{c c c}
-te^{ik_{x,t}a} & 0 & \cdots \\
0 & 0 & \\
\vdots &  & \ddots
\end{array}
\right], \nonumber \\
\bf{\Sigma_{b}}=\left[
\begin{array}{c c c}
\ddots & & \vdots \\
 & 0 & 0 \\
\cdots & 0 & -te^{ik_{x,b}a}
\end{array}
\right], \nonumber \\
\end{eqnarray}
, where $k_{x,t}$ and $k_{x,b}$ are longitudinal electron wave vectors inside top and bottom electrodes, respectively, given as
\begin{eqnarray}
k_{x,t}=\frac{\cos^{-1}\left\lbrace1-\frac{E-U\left(x_{1}\right)-E_{\perp}\left(\theta\right)}{2t}\right\rbrace}{a}, \\
k_{x,b}=\frac{\cos^{-1}\left\lbrace1-\frac{E-U\left(x_{N}\right)-E_{\perp}\left(\theta\right)}{2t}\right\rbrace}{a}.
\end{eqnarray}
In addition to TER in FTJs, the same approach can also be applied to other problems such as spin injection from a ferromagnet into a semiconductor or a metal \cite{6834821,6818391}, as long as the energy band diagram is known.

\bibliography{apssamp}

\end{document}